\documentclass[twocolumn]{aastex631}

\usepackage{amsmath}
\usepackage{graphicx}
\usepackage{units}
\usepackage{comment}
\usepackage{placeins} 

\graphicspath{{./}{Figures/}}
\usepackage{hyperref}

\begin{document}

\title{High-Resolution DKIST/ViSP Observation Reveals Small-scale Photospheric Magnetic Flux beneath a Coronal Hole Region.}

\author{Khagendra Katuwal}
\affiliation{Department of Astronomy, New Mexico State University, Las Cruces, NM 88003, USA}

\author{R. T. James McAteer}
\affiliation{Department of Astronomy, New Mexico State University, Las Cruces, NM 88003, USA}

\author{Rahul Yadav}
\affiliation{Laboratory for Atmospheric and Space Physics, University of Colorado Boulder, Boulder, CO 80303, USA}

\begin{abstract}
The magnetic unipolarity of photospheric regions underlying coronal holes is well established at spatial resolutions of $\sim 1^{\prime\prime}$. However, the extent to which small-scale magnetic structures affect the degree of magnetic unipolarity remains uncertain. To address this question, we present co-aligned observations from the Helioseismic and Magnetic Imager (HMI) onboard the \textit{Solar Dynamics Observatory} (\textit{SDO}) and the Visible Spectro-Polarimeter (ViSP) at the Daniel K. Inouye Solar Telescope (DKIST). HMI observes the \ion{Fe}{1} 6173~\AA\ line, while ViSP observes the \ion{Fe}{1} 6302.5~\AA\ line; both lines form at similar photospheric heights. We derive the line-of-sight magnetic field ($B_{\mathrm{los}}$) from the ViSP spectropolarimetric data using a Milne--Eddington (ME) inversion and compare it with the HMI $B_{\mathrm{los}}$ map derived from vector magnetograms over a common coronal-hole region. ViSP detects smaller magnetic structures with stronger field strength than HMI, with the absolute mean field ($|\langle B_{\mathrm{los}} \rangle|$) and the mean absolute field ($\langle |B_{\mathrm{los}}| \rangle$) enhanced by factors of 1.30 and 1.37, respectively. HMI measures approximately $28\%$ less unsigned magnetic flux and $24\%$ less net signed magnetic flux than ViSP within the common field of view. A comparison of the degraded ViSP $B_{\mathrm{los}}$ map with the HMI $B_{\mathrm{los}}$ map indicates that differences in effective spatial resolution and the associated spatial averaging contribute to the observed differences between the measurements. Despite the additional small-scale flux detected by ViSP, the degree of magnetic unipolarity remains similar, as indicated by the magnetic flux imbalance values of $69\%$ for HMI and $65\%$ for ViSP. These results indicate that the additional small-scale magnetic flux detected by ViSP does not substantially alter the degree of magnetic unipolarity within the observed coronal-hole region. However, because ViSP has a limited field of view, future observations covering a full coronal hole are needed to determine whether the same result applies to an entire coronal hole.
\end{abstract}


\FloatBarrier

\section{Introduction}\label{sec:intro}
\setcounter{figure}{0}

\begin{figure*}[!t]
  \centering
  \includegraphics[width=\textwidth]{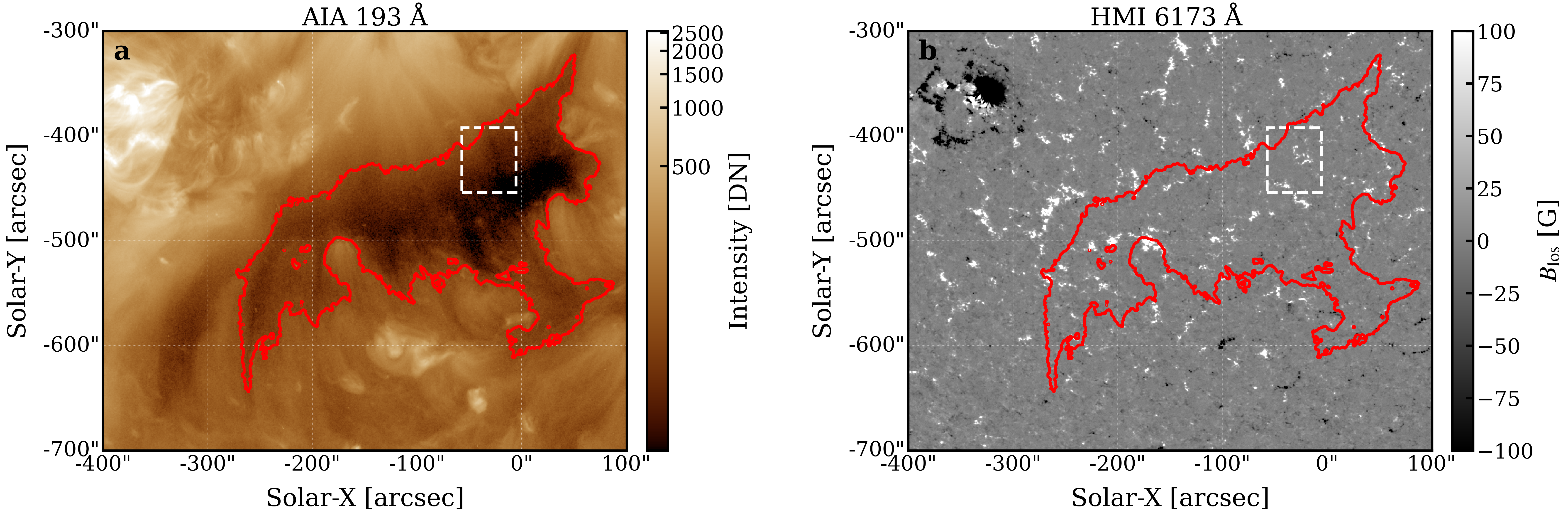}
  \caption{Panel `a': AIA 193~\AA\ intensity map showing the
  coronal hole as a region of reduced EUV emission. Panel `b':
  Co-temporal HMI magnetogram (720~s cadence) of the same region.
  The red contours in both panels outline the coronal-hole boundary.
  The dashed white box marks the high-resolution field of view observed
  by Visible spectro-polarimetry (ViSP) in the \ion{Fe}{1} 6302.5~\AA\ spectral line, which is the
  region analyzed in this study. The observations were obtained on
  2024 April 24 from 19:42 to 20:29~UT.}
  \label{fig:AIA_HMI_overview}
\end{figure*}

Coronal holes exhibit distinct observational signatures across different layers of the solar atmosphere. In the upper atmosphere, including the corona and upper transition region, they display reduced emission in extreme-ultraviolet (EUV) and soft X-ray (SXR) wavelengths \citep{1976Ap&SS..39...75V, CRANMER20023, 2022ApJ...932..115S}. This lower intensity is caused by plasma densities and temperatures that are significantly lower than those of the surrounding quiet Sun (QS) and active regions (AR). The reduced electron density strongly contributes to the lower EUV and X-ray intensity of coronal holes, since optically thin coronal emission scales approximately as \(n_e^2 G(T)\) \citep{2018LRSP...15....5D}, where \(n_e\) is the electron density and \(G(T)\) is the temperature-dependent contribution function. Thus, even a modest decrease in density can produce a substantial reduction in emission. The temperature difference between coronal holes and the quiet Sun is comparatively smaller, but it is sufficient to shift the plasma away from the temperature range in which commonly observed coronal EUV lines are formed, further reducing the observed intensity. EUV bands centered at 193~\AA, 171~\AA, and 211~\AA\ provide the highest contrast between coronal holes and non-coronal hole regions \citep{2024ApJS..271....6R, hofmeister2019photospheric} making these wavelengths most useful for identifying coronal holes. 

In the chromosphere, coronal holes are observed as features of enhanced intensity in \ion{He}{1} 10830~\AA\ filtergrams, resulting from reduced chromospheric absorption of the underlying continuum emission \citep{1996SoPh..163...79B}. However, their weak contrast with the surrounding chromosphere makes them difficult to identify reliably \citep{1983SoPh...87...47K}. In the photosphere, coronal holes do not exhibit a distinct intensity signature and therefore cannot be directly identified in continuum intensity observations \citep{1979SSRv...23..139H}.

Coronal holes are characterized by a predominantly unipolar magnetic field configuration \citep[e.g.][]{1979SSRv...23..139H, 1982SoPh...79..149H,2023AAS...24221404K,2023SPD....5420402K}. In them, magnetic field lines originate below the solar surface and extend outward through the solar atmosphere into interplanetary space. An open magnetic field topology enables the solar plasma to escape more efficiently from the solar atmosphere at a speed of approximately $400\text{--}800~\mathrm{km~s^{-1}}$ \citep{2006SSRv..124...51S}. As a result, coronal holes are widely recognized as the primary sources of the fast solar wind \citep{1973SoPh...29..505K, 1976SoPh...46..303N, wang1990solar, mccomas2007understanding}. Coronal holes are also the primary source of open magnetic flux in the upper solar atmosphere and play a crucial role in shaping the heliospheric magnetic field \citep{1996Sci...271..464W, 2009SSRv..144..383W}. Direct measurements of these coronal magnetic fields remain observationally challenging due to the low plasma density, high temperature, and weak magnetic-field strength of the solar corona \citep{Schad_2023,2024SciA...10.1604S,2024ApJ...963...69B}. As an alternative approach, magnetic field extrapolation techniques are often employed to infer the coronal and heliospheric magnetic fields from photospheric measurements \citep{https://doi.org/10.1029/2007JA012432}. These extrapolations use photospheric magnetograms as the lower boundary condition. Potential-field extrapolations use the normal or radial magnetic-field component of the photospheric magnetic field, which can be estimated from line-of-sight magnetograms \citep{1969SoPh....9..131A}, whereas nonlinear force-free field (NLFFF) extrapolations typically use all three components of the photospheric magnetic field vector \citep{2021LRSP...18....1W}. The accuracy and reliability of such extrapolations therefore critically depend on the accuracy and consistency of the underlying photospheric magnetic field measurements. 

In this context, the high spatial and spectral resolution of the Visible Spectro-Polarimeter (ViSP) \citep{2022SoPh..297...22D} detects small-scale photospheric magnetic structures that may remain unresolved in lower -resolution observations, providing better input for magnetic-field extrapolations. To date, magnetic-field extrapolations have been primarily used on photospheric magnetograms from instruments such as the Michelson Doppler Imager \citep[MDI; pixel scale $\approx 2\arcsec$, effective resolution $\approx 4\arcsec$;][]{1995SoPh..162..129S}, the Helioseismic and Magnetic Imager \citep[HMI; pixel scale $\approx 0.5\arcsec$, effective resolution $\approx 1\arcsec$;][]{2012SoPh..275..207S}, and the Global Oscillation Network Group \citep[GONG; spatial resolution $\approx 2.5\arcsec\text{--}5\arcsec$;][]{1996Sci...272.1284H}. Among currently available full-disk instruments, HMI provides the highest spatial resolution. However, even this resolution remains insufficient to resolve the smallest relevant magnetic structure. Numerical simulations suggest that individual photospheric magnetic structures may be as small as 0.05\arcsec, corresponding to approximately 35~km \textbf{\citep{2010ApJ...714.1606P,2014ApJ...789..132R}}. To reliably infer photospheric magnetic fields, an instrument must have sufficient spatial resolution to resolve magnetic structures at fundamental physical scales. Because these magnetic fields are commonly used as lower boundary conditions for coronal and heliospheric magnetic-field models, resolution-related spatial averaging can introduce uncertainties into the modeled fields.

The identification of coronal holes typically begins with visual inspection of coronal images, where they are usually identified as contiguous dark regions. However, not all dark coronal regions are coronal holes, since filaments \citep{2014CEAB...38...95R} and coronal voids \citep{2023A&A...678A.196N} can also exhibit reduced emission. Magnetic unipolarity is therefore a key follow-up criterion for validating dark regions as coronal holes, since filaments and coronal voids generally exhibit bipolar magnetic field distributions. This distinction in polarity is well established at a spatial resolution of approximately 1\arcsec\ using HMI photospheric magnetograms. 
The key question is whether ViSP's higher spatial resolution alters the inferred magnetic properties of the coronal-hole region. These properties include the total unsigned flux ($\Phi_u$), total signed flux ($\Phi_s$) absolute means field ($|\langle B_{\mathrm{los}}\rangle|$), mean absolute field ($\langle|B_{\mathrm{los}}|\rangle$), the polarity ratio ($\Phi_r$), the skewness ($S$), the flux imbalance ($\Phi_{\mathrm{imb}}$). Their mathematical definitions are presented in the second column of Table~\ref{tab:flux_properties}. Comparing these quantities between ViSP $B_{\mathrm{los}}$ and HMI $B_{\mathrm{los}}$ allows us to determine how small-scale magnetic features that are unresolved or spatially averaged in HMI affect the inferred field strength, the $B_{\mathrm{los}}$ distribution, and the magnetic unipolarity of the region. 

In this work, high-resolution spectropolarimetric observations obtained with ViSP and retrieved from the Daniel K. Inouye Solar Telescope (DKIST; \citealt{2020SoPh..295..172R}) Data Center were used to investigate the photospheric magnetic field. ViSP measures wavelength-dependent Stokes profiles, which contain information about the magnetic field in the solar atmosphere. Because the magnetic field cannot be measured directly, it is inferred from its signatures in polarized spectral lines using diagnostic models, techniques, and assumptions. We infer $B_{\mathrm{los}}$ from the \ion{Fe}{1} 6302.5~\AA\ line using a Milne--Eddington (ME) inversion and compare the resulting magnetic-field properties with those derived from the corresponding HMI vector magnetograms.

The paper is organized as follows: Section~\ref{sec:data} describes the observational data and setup; Section~\ref{sec:method} outlines the methodology; Section~\ref{sec:Result} presents the results and analysis; and Section~\ref{sec:Con} summarizes the main conclusions.

\section{Datasets}\label{sec:data}

This study combines space-based and ground-based observations to investigate $B_{\mathrm{los}}$ in a common photospheric region beneath a coronal hole. The space-based observations were obtained with the Atmospheric Imaging Assembly (AIA; \citep{2012SoPh..275...17L}) and the Helioseismic and Magnetic Imager (HMI; \citep{2012SoPh..275..207S}) onboard the Solar Dynamics Observatory (SDO; \citep{2012SoPh..275....3P}). The AIA data were used to identify the coronal hole in the solar corona, while the HMI data were selected to investigate the magnetic properties of the underlying photospheric region. Ground-based observations acquired with ViSP at DKIST were also used to study the photospheric magnetic properties at higher spatial resolution. The coronal hole region analyzed in this study was located at approximately $-30^\circ$ solar latitude and near the central meridian. The observations were obtained on 2024~April~24 between 19{:}42~UT and 20{:}29~UT using both DKIST and SDO.

\subsection{SDO}
SDO carries three instruments. Data from two instruments, AIA and HMI are used in this study. AIA provides full-disk EUV images with a plate scale of $0^{\prime\prime}\!.6$\,pixels$^{-1}$, an effective spatial resolution of $1^{\prime\prime}\!.5$, and a cadence of $12$\ s \citep{2012SoPh..275...17L}. The AIA 193\,\AA\ channel is used to identify the coronal hole. Figure~\ref{fig:AIA_HMI_overview}a shows an AIA 193\,\AA\ intensity map of the solar corona, in which the coronal hole appears as a region of reduced intensity. The red contour outlines the coronal-hole boundary, identified as described in Section~\ref{sec:method}. HMI provides full-disk $B_{\mathrm{los}}$ measurements every $\sim45$\,s, with a plate scale of $0^{\prime\prime}\!.5\, 
\mathrm{pixel}^{-1}$ and an effective spatial resolution of $\sim 0^{\prime\prime}\!.91$ in the photosphere \citep{2012SoPh..275..207S}. These measurements were derived from spectro-polarimetric observations of the \ion{Fe}{1} 6173~\AA\ spectral line. HMI also provides full-disk vector magnetograms, derived from Stokes inversions of the same \ion{Fe}{1} 6173~\AA\ spectral line, at a cadence of  $\sim720$\ s \citep{2014SoPh..289.3483H}. The $B_{\mathrm{los}}$ derived from the HMI vector magnetograms is used in this study because it provides a higher signal-to-noise ratio than the HMI 45 s line-of-sight magnetograms. Figure~\ref{fig:AIA_HMI_overview}b shows the HMI $B_{\mathrm{los}}$ map of the photospheric magnetic field beneath the coronal hole region shown in Figure~\ref{fig:AIA_HMI_overview}a. The red contour is the same coronal hole boundary shown in Figure~\ref{fig:AIA_HMI_overview}a.
Science-ready Level~1.5 AIA and HMI data were obtained via the Virtual Solar Observatory\footnote{\url{https://virtualsolar.org}} using \texttt{SunPy} \citep{sunpy_community2020}. An example of the data-acquisition procedure can be found at the Zenodo link \citep{katuwal_2025_16521356}

\subsection{DKIST}
\begin{figure*}[!t]
  \centering
  \includegraphics[width=\textwidth]{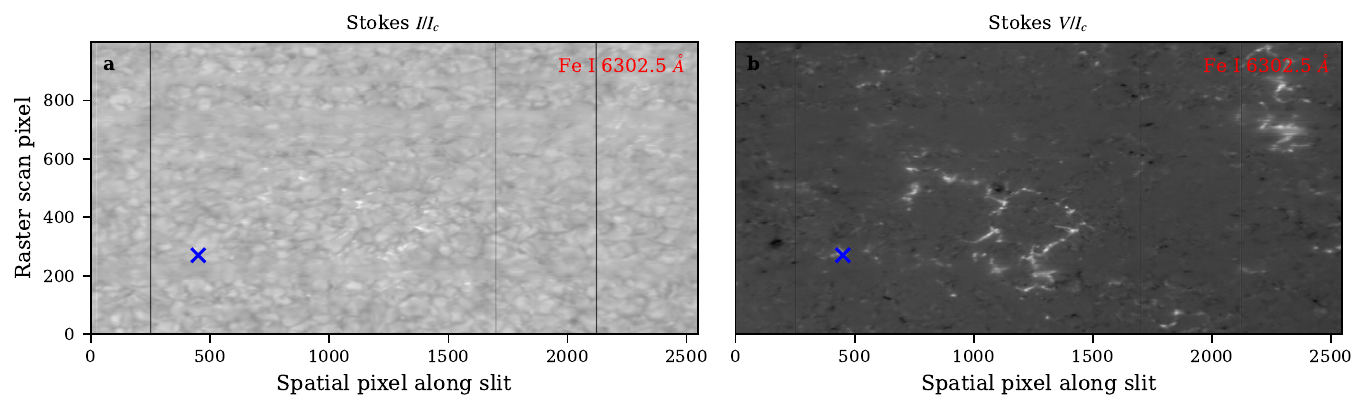}
  \caption{Spatial maps of the \ion{Fe}{1} 6302.5~\AA\ line obtained from the ViSP raster corresponding to the region outlined by the white dotted box in Figure~\ref{fig:AIA_HMI_overview}. Panel `a' shows the Stokes $I/I_c$ intensity, while panel `b' shows the corresponding Stokes $V/I_c$ signal. Both quantities were averaged over wavelength indices 590--607 ( corresponding to the wavelength range 6302.392--6302.595~\AA) around the line core (\ion{Fe}{1} 6302.5~\AA). The horizontal axis represents the spatial position along the slit, while the vertical axis represents the raster-scan step. The blue cross marks the selected location at slit pixel 450 and raster step 270, which was used to examine the effect of the crosstalk correction on representative Stokes profiles. See Appendix~\ref{sec:appendix} for details.}
\label{fig:FeI6302_Stokes_maps}
\end{figure*}

ViSP at DKIST is a slit-scanning spectro-polarimeter designed to provide high spatial and spectral resolution observations. ViSP records the full Stokes vector $(I, Q, U, V)$, where $I$ is the intensity, $Q$ and $U$ are linear polarizations, and $V$ is the circular polarization, over the wavelength range 3800--9000\,\AA. The instrument can simultaneously observe up to three spectral regions using separate detectors.

In this observation, spectral windows containing the \ion{Fe}{1} 6301.5~\AA\ and \ion{Fe}{1} 6302.5~\AA\ lines, the \ion{Na}{1} D$_1$ 5896~\AA\ line, and the \ion{Ca}{2} 8542~\AA\  line were recorded. These lines probe different layers of the solar atmosphere.  The \ion{Fe}{1} 6301.5~\AA\ and \ion{Fe}{1} 6302.5~\AA\ lines form in the photosphere, the \ion{Na}{1} D$_1$ 5896~\AA\ samples the upper photosphere to the lower chromosphere, and the \ion{Ca}{2} 8542~\AA\ primarily forms in the chromosphere. Since this work focuses on the $B_{\mathrm{los}}$ at photospheric heights, we analyzed only the \ion{Fe}{1} 6302.5~\AA\ spectral line.  The inferred ViSP $B_{\mathrm{los}}$ is compared with the HMI $B_{\mathrm{los}}$ derived from an HMI vector magnetogram, as both diagnostics probe similar photospheric layers \citep{2017ApJ...851..111S}.

The white dotted box shown in both panels of Figure~\ref{fig:AIA_HMI_overview} indicates the ViSP field of view. This observation is obtained under Proposal ID \texttt{pid\_2\_28}. The data corresponds to the Level~1 product \texttt{L1-WKCSV}, processed with the \texttt{l0\_to\_l1\_visp} calibration workflow version \texttt{5.4.0}\footnote{\url{https://docs.dkist.nso.edu/projects/visp/en/v5.4.0/l0_to_l1_visp.html}}. The dataset consists of a 1000-step spectro-polarimetric raster with four Stokes parameters $(I,Q,U,V)$. The observed spectral window spans 6296.49--6307.29\,\AA, covering the 
\ion{Fe}{1} 6301.5~\AA\ and \ion{Fe}{1} 6302.5~\AA\ lines, with a spectral sampling of 0.012\,\AA\,pixel$^{-1}$. The spatial sampling is $0^{\prime\prime}\!.024$
along the slit and $0^{\prime\prime}\!.052$ in the raster direction, with a slit width of $0^{\prime\prime}\!.054$. With 1000 raster steps and 2547 spatial pixels along the slit, the observations cover a field of view of approximately $62'' \times 52''$. The ViSP field of view spans helioprojective coordinates from $(-5.40\arcsec,-391.75\arcsec)$ to $(-57.16\arcsec,-453.51\arcsec)$ relative to disk center, with a center located at $(X,Y)=(-31\farcs28,-422\farcs63)$. The average temporal cadence between successive raster steps is 2.82\,s. 
The Level~1 data values are normalized relative to the average solar disk-center intensity following the standard ViSP calibration procedure. These observations target the photospheric magnetic field underlying a coronal hole region. They are used to investigate the small-scale magnetic flux distribution and its contribution to the open magnetic field. The high spatial, spectral, and polarimetric precision of ViSP enables detailed measurements of fine-scale magnetic elements that remain unresolved in lower-resolution observations. 

\subsection{Data Reduction}
The Level~1 data used in this study are publicly available through the National Solar Observatory (NSO)/DKIST Archive\footnote{\url{https://dkist.data.nso.edu/datasetview/PMOEUM}}, and the basic calibration steps have already been applied. These include dark and background subtraction, gain calibration, geometric-distortion correction, polarimetric demodulation, beam combination, and correction for telescope-induced polarization. Details of the ViSP Level~1 calibration pipeline are provided in the DKIST documentation.\footnote{\url{https://docs.dkist.nso.edu/projects/visp/en/v5.4.0/science_calibration.html}} The continuum-normalized Stokes intensity ($I/I_c$), and circular-polarization ($V/I_c$), maps from this dataset are shown in Figure~\ref{fig:FeI6302_Stokes_maps}a and Figure~\ref{fig:FeI6302_Stokes_maps}b, respectively. The $I/I_c$ map shows spatial variations in the continuum-normalized \ion{Fe}{1} 6302.5~\AA\ line intensity, whereas the $V/I_c$ map shows the circular-polarization signal produced primarily by the longitudinal Zeeman effect. It is therefore sensitive to the strength and polarity of $B_{\mathrm{los}}$. Our objective is to derive $B_{\mathrm{los}}$ from the observed circular polarization profiles. We need additional processing of these data before performing the inversions. This processing includes polarization crosstalk correction and wavelength calibration of the \ion{Fe}{1} 6302.5~\AA\ spectral line; further details are available at the Zenodo link \citep{katuwal_2026_20825239}. The blue crosses in both panels of Figure~\ref{fig:FeI6302_Stokes_maps} mark the selected location at raster step 270 and spatial pixel 450. This location is used to demonstrate the effect of the crosstalk correction. The correction procedure and its effect on the representative Stokes profiles are presented and discussed in detail in Appendix~\ref{sec:appendix}.
The absolute wavelength calibration is obtained by matching the mean quiet-Sun spectrum from the ViSP observations to the disk-center reference spectrum from the BASS~2000 solar atlas\footnote{\url{https://bass2000.obspm.fr/solar_spect.php}}, as described in \cite{2024ApJ...973L..10Y}. The data are then rebinned along the slit to improve the signal-to-noise ratio and produce approximately square spatial pixels. The resulting processed Stokes profiles are used as input to the ME inversions. The data preprocessing code is available on Zenodo link \citep{katuwal_2026_20825239}
\section{Methodology}\label{sec:method}
The observational target is a coronal hole region. Because coronal holes are difficult to identify directly at photospheric heights, we use the AIA 193~$\mathring{\text{A}}$ channel as a reference, as shown in Figure~\ref{fig:AIA_HMI_overview}a. This wavelength provides a strong contrast between coronal holes and the surrounding corona \citep{2024ApJS..271....6R} and is commonly used to identify coronal hole boundaries. The boundary is determined using an intensity threshold of 100~DN, following \citep{katuwal2026unipolarity}, and is indicated by the red contour outlining the region of reduced emission. The red contour is then overlaid on the $B_{\mathrm{los}}$ map derived from the HMI vector magnetogram to investigate the magnetic properties of the photospheric region underlying the coronal hole, as shown in Figure~\ref{fig:AIA_HMI_overview}b. HMI data products are ready to use for magnetic-field analysis. However, ViSP data require further processing and inversion to obtain $B_{\mathrm{los}}$.

\subsection{Magnetic Field Inversions}\label{subsec:inversions}

The ViSP instrument does not directly provide $B_{\mathrm{los}}$. Instead, it records the full Stokes profiles $(I,Q,U,V)$, which requires further postprocessing to infer the magnetic field vector. We derive $B_{\mathrm{los}}$ from the \ion{Fe}{1} 6302.5~$\mathring{\text{A}}$ spectral line using ME inversion of the Stokes profiles using the SPIN inversion code \citep{2017SoPh..292..105Y}, which solves the radiative transfer equation under a ME atmospheric framework. This model assumes that all physical parameters remain constant with height, except for the source function, which varies linearly with optical depth.  After retrieving $B_{\mathrm{los}}$, the resulting ViSP $B_{\mathrm{los}}$ map is co-aligned with the HMI $B_{\mathrm{los}}$ map. For reference, the HMI magnetogram has a noise level of approximately $5$--$10~\mathrm{G}$\citep{2012SoPh..275..207S,2012SoPh..275..229S,2016SoPh..291.1887C}, while the ViSP measurements have an estimated noise level of a few gauss. These values provide an approximate indication
of the weakest magnetic fields detectable by the two instruments.
Within the identified coronal-hole region, we calculate magnetic flux properties, flux-enhancement ratios, distribution properties, and the fractions of unsigned and signed flux that HMI misses. These quantities are then compared between the two instruments. A complete summary of the comparison of the derived magnetic characteristics is presented in Table~\ref{tab:flux_properties}.

\subsection{Spatial Co-alignment of HMI and ViSP}\label{subsec:alignment}

To co-align the high-resolution ViSP \ion{Fe}{1} 6302.5~\AA\ observations with the HMI magnetograms, we constructed a reference $B_{\mathrm{los}}$ map from the HMI 720~s vector data product, \texttt{hmi.b\_720s}. Using the native helioprojective World Coordinate System (WCS) preserved in the HMI FITS header, the line-of-sight magnetic field was computed as
\begin{equation}
B_{\mathrm{los}} = B \cos \gamma ,
\end{equation}
where $B$ is the magnetic-field strength and $\gamma$ is the field inclination. The corresponding ViSP $B_{\mathrm{los}}$ map was derived from the Level-1 Stokes~$V$ FITS file, with the spatial plate scales along the slit and raster directions extracted from the DKIST header keywords \texttt{CDELT1} and \texttt{CDELT2}.

\begin{figure*}[!t]
  \centering
  \includegraphics[width=\textwidth]{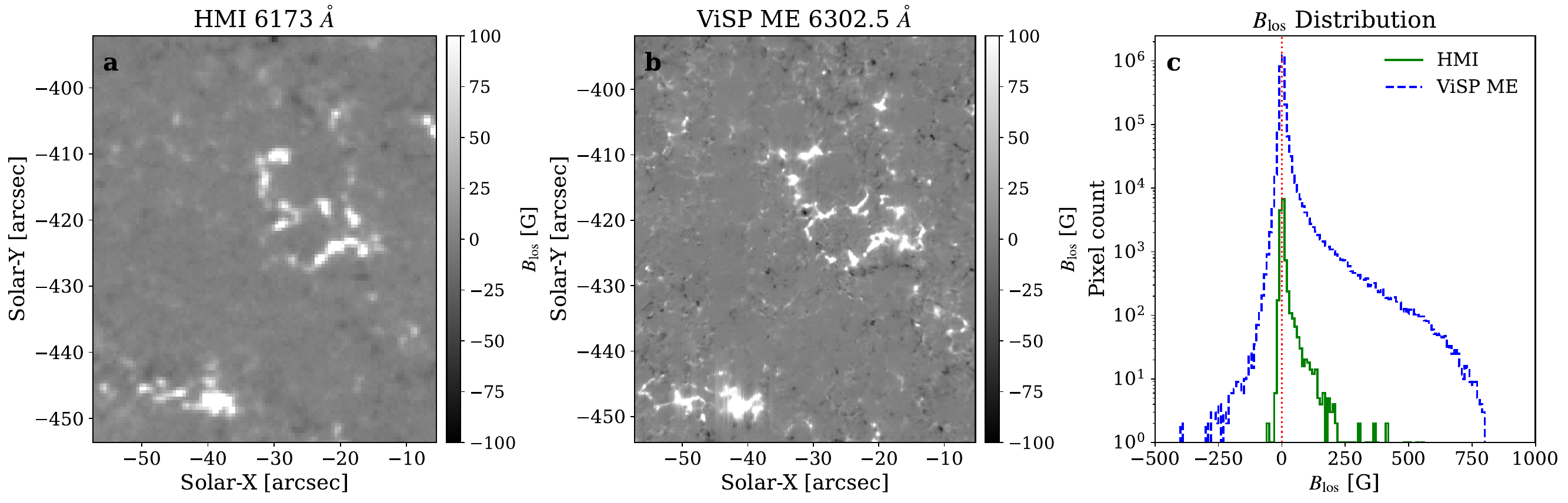}
  \caption{Panels `a' and `b' show co-aligned images of the same photospheric field of view observed by HMI and ViSP, respectively. This region corresponds to the area marked by the dotted rectangle in Figure~\ref{fig:AIA_HMI_overview}. Panel `c' shows histograms comparing the $B_{\mathrm{los}}$ distributions in panels `a' and `b' and the vertical dashed red line indicates \(B_{\mathrm{los}}=0\) G.}
  \label{fig:Figure5}
\end{figure*}

The ViSP field of view, indicated by the white dotted box in Figure~\ref{fig:AIA_HMI_overview}, was defined using two helioprojective coordinate pairs, $(-5.40^{\prime\prime}, -391.75^{\prime\prime})$ and $(-57.16^{\prime\prime}, -453.51^{\prime\prime})$, measured relative to the center of the solar disk. These coordinates were converted into the native \texttt{Helioprojective} coordinate frame using the \texttt{SunPy} software package to extract the corresponding spatial submap from the full-disk HMI data. To account for the ViSP slit-scanning geometry, the ViSP raster was rotated by $90^\circ$ and flipped horizontally to align its spatial axes with helioprojective longitude and latitude. This procedure ensured that the resolved magnetic polarities were physically consistent with the reference HMI $B_{\mathrm{los}}$ map.  The final alignment was verified by plotting the ViSP
$B_{\mathrm{los}}$ (Figure~\ref{fig:Figure5}b) and HMI
$B_{\mathrm{los}}$ (Figure~\ref{fig:Figure5}a) images side by side
on the same helioprojective arcsecond scale within the shared WCS
framework. The two panels cover an identical field of view. 

\section{Results and Analysis}\label{sec:Result}
\subsection{Visual Comparison of the HMI and ViSP Magnetograms}
Figure~\ref{fig:Figure5}a and Figure~\ref{fig:Figure5}b show the co-aligned HMI and ViSP magnetograms, respectively. The main positive-polarity features near the center and the lower-left edge appear at the same locations and have similar shapes in both maps. The ViSP $B_{\mathrm{los}}$ map (Figure~\ref{fig:Figure5}b) displays finer spatial features and resolves additional small-scale magnetic features, particularly within the weak background field and around the main magnetic concentrations. Most of the field of view contains weak $B_{\mathrm{los}}$ values near zero, together with small patches of both polarities. The positive-polarity features are generally larger and more elongated, whereas the negative-polarity features are smaller and more widely scattered. In the HMI $B_{\mathrm{los}}$ (Figure~\ref{fig:Figure5}a), the positive-polarity features appear more blurred, and many of the small negative-polarity features are less clearly visible. Overall, the region contains weak, mixed-polarity photospheric fields and no strong sunspot-like magnetic concentrations. This visual comparison provides context for the quantitative analysis in which we examine whether the similarities and differences seen in the two maps are also reflected in their magnetic field distributions and derived magnetic properties.

\subsection{Distribution of \texorpdfstring{$B_{\mathrm{los}}$}{B-los} }
Figure~\ref{fig:Figure5}c compares the \(B_{\mathrm{los}}\) distributions measured by HMI (green curve) and ViSP (blue curve) over the same field of view. Both curves peak strongly near \(B_{\mathrm{los}}\approx 0\)~G, indicating that relatively weak magnetic fields occupy most of the region. To quantify this concentration, we calculate the fraction of pixels satisfying \(|B_{\rm los}|<|B_{\rm los}|_{\max}/10\) for each distribution and determine the unsigned magnetic flux contributed by those pixels relative to the total unsigned flux. For ViSP, \(|B_{\rm los}|_{\max}/10=79.8\)~G, and 98.3\% of the pixels fall within this range, contributing approximately 68.2\% of the total unsigned magnetic flux. For HMI, \(|B_{\rm los}|_{\max}/10=55.5\)~G, and 98.1\% of the pixels fall within this range, contributing approximately 68.0\% of the total unsigned magnetic flux. The HMI \(B_{\mathrm{los}}\) distribution is relatively narrow, whereas the ViSP distribution is substantially broader and spans a wider range of field strengths. The ViSP distribution extends to approximately \(800\)~G on the positive side and \(-297\)~G on the negative side, compared with approximately \(555\)~G and \(-57\)~G, respectively, for HMI.
An isolated ViSP value near \(-400\)~G is treated as an artifact and is therefore not considered representative of the distribution. This value is spatially isolated, and the
corresponding Stokes \(V\) signal does not show a
sufficiently strong circular-polarization signal to support a magnetic field of approximately \(-400\)~G. Including or excluding this outlier does not significantly affect the calculated statistics. For HMI, however, the high-field pixels with \(B_{\mathrm{los}}>250\)~G occur mainly in a compact group of neighboring pixels, with field strengths extending up to approximately \(555\)~G. Although
some of these values appear in histogram bins
containing only one or two pixels, they are
associated with a coherent magnetic concentration
and are therefore not considered artifacts. Both \(B_{\mathrm{los}}\) distributions are positively asymmetric, consistent with the dominance of positive-polarity magnetic fields in the coronal-hole region. The broader ViSP distribution indicates that ViSP detects stronger small-scale magnetic concentrations that are spatially averaged in HMI, supporting the visual evidence that ViSP reveals additional small-scale magnetic flux that is less clearly resolved in the HMI map.
\begin{table*}[htbp]
\centering

\caption{Magnetic Flux Properties: HMI vs.\ ViSP}
\label{tab:flux_properties}

\renewcommand{\arraystretch}{1.25}
\setlength{\tabcolsep}{5pt}

\makebox[\textwidth][c]{%
\resizebox{1.12\textwidth}{!}{%
\begin{tabular}{|l|l|c|c|}
\hline

\textbf{Property}
&
\textbf{Definition}
&
\textbf{HMI}
&
\textbf{ViSP ME Inversion}
\\
\hline

\multicolumn{4}{|c|}{
\textit{Magnetic Flux Properties}
}
\\
\hline

Total unsigned flux, $\Phi_u$ [Mx]
&
$\displaystyle
\Phi_u=\sum |B_{\mathrm{los}}|A$
&
$1.189\times10^{20}$
&
$1.656\times10^{20}$
\\
\hline

Total signed flux, $\Phi_s$ [Mx]
&
$\displaystyle
\Phi_s=\sum B_{\mathrm{los}}A$
&
$8.162\times10^{19}$
&
$1.073\times10^{20}$
\\
\hline

Positive flux, $\Phi^{+}$ [Mx]
&
$\displaystyle
\sum\limits_{B_{\mathrm{los}}>0}
B_{\mathrm{los}}A$
&
$1.003\times10^{20}$
&
$1.364\times10^{20}$
\\
\hline

Negative-flux magnitude, $|\Phi^{-}|$ [Mx]
&
$\displaystyle
\left|
\sum\limits_{B_{\mathrm{los}}<0}
B_{\mathrm{los}}A
\right|$
&
$1.866\times10^{19}$
&
$2.913\times10^{19}$
\\
\hline

Polarity-flux ratio, $\phi_r$
&
$\displaystyle
\frac{\Phi^{+}}{|\Phi^{-}|}$
&
$5.4$
&
$4.7$
\\
\hline

Absolute mean field [G]
&
$\displaystyle
\left|
\left\langle B_{\mathrm{los}}\right\rangle
\right|$
&
$4.9$
&
$6.3$
\\
\hline

Mean absolute field [G]
&
$\displaystyle
\left\langle |B_{\mathrm{los}}| \right\rangle$
&
$7.1$
&
$9.7$
\\
\hline

\multicolumn{4}{|c|}{
\textit{Flux Enhancement (ViSP/HMI)}
}
\\
\hline

Total unsigned-flux enhancement
&
$\displaystyle
\frac{\Phi_u^{\mathrm{ViSP}}}
     {\Phi_u^{\mathrm{HMI}}}$
&
\textemdash
&
$1.4$
\\
\hline

Positive-flux enhancement
&
$\displaystyle
\frac{\Phi_{\mathrm{ViSP}}^{+}}
     {\Phi_{\mathrm{HMI}}^{+}}$
&
\textemdash
&
$1.4$
\\
\hline

Negative-flux enhancement
&
$\displaystyle
\frac{|\Phi_{\mathrm{ViSP}}^{-}|}
     {|\Phi_{\mathrm{HMI}}^{-}|}$
&
\textemdash
&
$1.6$
\\
\hline

\multicolumn{4}{|c|}{
\textit{Distribution Properties}
}
\\
\hline

Flux imbalance, $\Phi_{\mathrm{imb}}$ [\%]
&
$\displaystyle
100\,\frac{|\Phi_s|}{\Phi_u}$
&
$69$
&
$65$
\\
\hline

Skewness, $S$
&
$\displaystyle
\frac{
3\left[
\left\langle B_{\mathrm{los}}\right\rangle
-
\operatorname{median}(B_{\mathrm{los}})
\right]
}{
\sigma_{B_{\mathrm{los}}}
}$
&
$0.50$
&
$0.46$
\\
\hline

\multicolumn{4}{|c|}{
\textit{Missing Flux in HMI Relative to ViSP}
}
\\
\hline

Missing unsigned flux, $\Phi_{m,\mathrm{unsigned}}$ [\%]
&
$\displaystyle
\left(
1-\frac{\Phi_u^{\mathrm{HMI}}}
        {\Phi_u^{\mathrm{ViSP}}}
\right)\times100$
&
\textemdash
&
$28$
\\
\hline

Missing signed flux, $\Phi_{m,\mathrm{signed}}$ [\%]
&
$\displaystyle
\left(
1-\frac{|\Phi_s^{\mathrm{HMI}}|}
        {|\Phi_s^{\mathrm{ViSP}}|}
\right)\times100$
&
\textemdash
&
$24$
\\
\hline

\end{tabular}%
}%
}

\vspace{5pt}

\makebox[\textwidth][c]{%
\parbox{1.12\textwidth}{%
\footnotesize
\setlength{\parindent}{0pt}%
\setlength{\leftskip}{0pt}%
\setlength{\rightskip}{0pt}%
\setlength{\parfillskip}{0pt plus 1fil}%
\noindent
Magnetic flux properties derived from the HMI $B_{\mathrm{los}}$
and ViSP $B_{\mathrm{los}}$ maps over the common field of view
within the coronal hole. The listed quantities include the total unsigned flux $\Phi_u$, total signed flux $\Phi_s$, positive flux $\Phi^{+}$,
negative-flux magnitude $|\Phi^{-}|$, polarity-flux ratio $\phi_r=\Phi^{+}/|\Phi^{-}|$, absolute mean field $|\langle B_{\mathrm{los}}\rangle|$, and mean absolute field $\langle|B_{\mathrm{los}}|\rangle$. The table also presents the ViSP-to-HMI flux-enhancement ratios, flux imbalance $\Phi_{\mathrm{imb}}$,
skewness $S$, and the percentages of missing unsigned and
signed magnetic flux, $\Phi_{m,\mathrm{unsigned}}$
and $\Phi_{m,\mathrm{signed}}$, respectively,
in HMI relative to the ViSP ME inversion.
}%
}

\end{table*}

\subsection{Magnetic unipolarity}
Magnetic unipolarity refers to the dominance of one magnetic polarity,
either positive or negative, in the region under study. The unsigned magnetic flux ($\Phi_u$) represents the total magnetic flux contributed by both positive and negative magnetic polarities, regardless of their signs. In contrast, the signed magnetic flux ($\Phi_s$) represents the net magnetic flux within a region, accounting for the contributions of both polarities. The unsigned magnetic flux is calculated as

\begin{equation}
\Phi_u = \sum_i |B_i|\,A_{\mathrm{pixel}},
\label{eq:unsigned_flux}
\end{equation}

and the signed magnetic flux is calculated as
\begin{equation}
\Phi_s = \sum_i B_i\,A_{\mathrm{pixel}},
\label{eq:signed_flux}
\end{equation}

where $B_i$ is the $B_{\mathrm{los}}$ in pixel $i$, and $A_{\mathrm{pixel}}$ is the physical area of a single pixel. The summation includes all valid pixels.

We characterize the magnetic-field distribution using the skewness ($S$) and quantify the degree of magnetic unipolarity using the flux imbalance ($\Phi_{\mathrm{imb}}$), defined as

\begin{equation}
S =
\frac{
3\left[
\langle B_{\mathrm{los}}\rangle
-
\mathrm{median}(B_{\mathrm{los}})
\right]
}{
\sigma_{B_{\mathrm{los}}}
},
\label{eq:skewness}
\end{equation}

and

\begin{equation}
\Phi_{\mathrm{imb}}
=\left(\frac{\Phi_s}{\Phi_u}\right) \times 100\%,
\label{eq:flux_imbalance}
\end{equation}

respectively, where $\sigma_{B_{\mathrm{los}}}$ is the standard deviation of the $B_{\mathrm{los}}$ distribution. Skewness measures the asymmetry in the shape of the distribution, whereas flux imbalance directly quantifies the predominance of one magnetic polarity over the other. These quantities therefore define key magnetic characteristics of coronal holes, which are generally associated with predominantly open and unipolar magnetic-field regions.

In our previous work \citep{katuwal2026unipolarity}, we partitioned the coronal hole into smaller subregions and investigated the degree of magnetic unipolarity in each region. Most of the analyzed subregions showed a clear dominance of one magnetic polarity. Figure~\ref{fig:Figure5}c compares the \(B_{\mathrm{los}}\) distributions measured by HMI and ViSP. Both datasets show positive skewness. The HMI map has a skewness of \(0.50\) and a magnetic flux imbalance of \(69\%\), while the ViSP map has a skewness of \(0.46\) and a magnetic flux imbalance of \(65\%\). These similar skewness values indicate comparable asymmetry in the two \(B_{\mathrm{los}}\) distributions, while the flux imbalance values indicate strong positive-polarity dominance in both datasets. The positive-to-negative flux ratios are likewise similar, with values of \(5.4\) for HMI and \(4.7\) for ViSP, confirming strong positive-polarity dominance in both measurements. The somewhat lower flux imbalance and flux ratio in ViSP indicate a relatively larger contribution from negative-polarity flux than in HMI.

The positive flux measured by ViSP is \(1.4\) times that measured by HMI, while the corresponding factor for negative flux is \(1.6\). Thus, ViSP measures \(40\%\) more positive flux and \(60\%\) more negative flux than HMI, indicating a substantial increase in detected magnetic flux despite the two instruments exhibiting similar overall polarity dominance. The larger relative increase in negative flux is consistent with ViSP detecting additional small-scale negative-polarity flux, leading to a modest reduction in the flux imbalance while preserving the region's overall positive-polarity dominance. However, high-resolution observations covering the full coronal-hole field of view are needed to confirm whether this behavior holds across the entire coronal hole.

\begin{figure*}[!t]
  \centering
  \includegraphics[width=\textwidth]{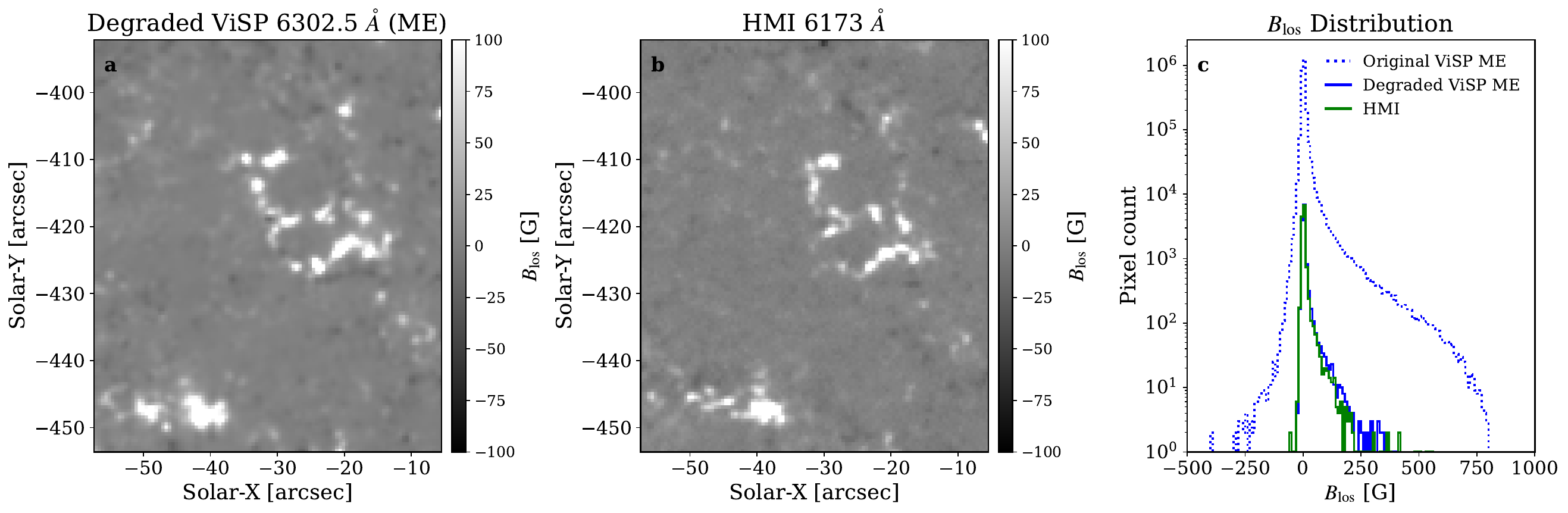}
  \caption{Comparison of $B_{\mathrm{los}}$ measurements from ViSP and HMI. Panel `a' shows the ViSP $B_{\mathrm{los}}$ map degraded to the HMI spatial resolution and pixel grid, while panel `b' shows the original HMI $B_{\mathrm{los}}$ map. Panel `c' compares the $B_{\mathrm{los}}$ distributions from the original ViSP $B_{\mathrm{los}}$ map (dotted blue curve), the degraded ViSP $B_{\mathrm{los}}$ map (solid blue curve), and HMI $B_{\mathrm{los}}$ (green curve).}
  \label{fig:Degraded_VISP_HMI_Histogram_APJ}
\end{figure*}

\subsection{Missing Flux}
 Using the definition from equation~\ref{eq:unsigned_flux}, we find that the unsigned magnetic flux measured by ViSP is
$\Phi_u^{\mathrm{ViSP}}=1.656\times10^{20}~\mathrm{Mx}$, whereas the corresponding HMI value is $\Phi_u^{\mathrm{HMI}}=1.189\times10^{20}~\mathrm{Mx}$.

The percentage difference in unsigned magnetic flux relative to ViSP is quantified as
\begin{equation}
\Phi_{m,\mathrm{unsigned}}
= \left(
1-\frac{\Phi_u^{\mathrm{HMI}}}
{\Phi_u^{\mathrm{ViSP}}}
\right)\times100\%.
\end{equation}

The difference in unsigned magnetic flux between the two measurements is $\Delta\Phi_u=4.670\times10^{19}~\mathrm{Mx}$,
indicating that HMI measures approximately $28\%$ less unsigned magnetic flux than ViSP. This difference includes contributions of $22\%$ from positive polarity and $6\%$ from negative polarity, relative to the total unsigned magnetic flux measured by ViSP. Thus, HMI measures less flux from both polarities, with the larger absolute difference associated with the dominant positive polarity.

Consistent with the larger unsigned flux measured by ViSP, the mean absolute line-of-sight magnetic field, $\langle |B_{\mathrm{los}}| \rangle$, is $9.7~\mathrm{G}$ for ViSP and $7.1~\mathrm{G}$ for HMI. Thus, the ViSP $\langle |B_{\mathrm{los}}| \rangle$ is larger than the HMI $\langle |B_{\mathrm{los}}| \rangle$ by a factor of $1.37$.

 Using the definition from equation~\ref{eq:signed_flux}, we find that the total signed magnetic flux measured by ViSP is
$\Phi_s^{\mathrm{ViSP}}=1.073\times10^{20}~\mathrm{Mx}$, whereas the corresponding HMI value is $\Phi_s^{\mathrm{HMI}}=8.162\times10^{19}~\mathrm{Mx}$. The percentage difference in signed magnetic flux magnitude relative to ViSP is quantified as
\begin{equation}
\Phi_{m,\mathrm{signed}}
=
\left(
1-
\frac{\left|\Phi_s^{\mathrm{HMI}}\right|}
{\left|\Phi_s^{\mathrm{ViSP}}\right|}
\right)
\times100\%.
\end{equation}

The difference in the signed-flux magnitude between the two measurements is
$\Delta\Phi_s=2.568\times10^{19}~\mathrm{Mx}$,
indicating that the magnitude of the signed magnetic flux measured by HMI is approximately $24\%$ lower than that measured by ViSP. Consistent with the larger signed flux measured by ViSP, the absolute mean line-of-sight magnetic field, $\left|\langle B_{\mathrm{los}}\rangle\right|$, is $6.3~\mathrm{G}$ for ViSP and $4.9~\mathrm{G}$ for HMI. Thus, the ViSP $\left|\langle B_{\mathrm{los}}\rangle\right|$ is larger than the HMI $\left|\langle B_{\mathrm{los}}\rangle\right|$ by a factor of $1.30$.
\subsection{Effect of Spatial Resolution on the ViSP--HMI Comparison}

Figure~\ref{fig:Degraded_VISP_HMI_Histogram_APJ}a shows the degraded ViSP \(B_{\mathrm{los}}\) map derived from panel `b' of Figure~\ref{fig:Figure5}, which presents the original ViSP \(B_{\mathrm{los}}\) map. The original ViSP \(B_{\mathrm{los}}\) map was convolved with a Gaussian kernel chosen to approximate the HMI spatial resolution and was then resampled onto the native HMI pixel grid. The resulting degraded ViSP \(B_{\mathrm{los}}\) map is compared with the original HMI \(B_{\mathrm{los}}\) map in Figure~\ref{fig:Degraded_VISP_HMI_Histogram_APJ}b. These two maps show broadly similar spatial distributions.

In Figure~\ref{fig:Degraded_VISP_HMI_Histogram_APJ}c, the dotted blue curve represents the original ViSP \(B_{\mathrm{los}}\) distribution, the solid blue curve represents the degraded ViSP \(B_{\mathrm{los}}\) distribution, and the green curve represents the HMI \(B_{\mathrm{los}}\) distribution. The degraded ViSP distribution is narrower than the original ViSP distribution and is more similar to the HMI distribution near the central peak and along the negative-field tail, while noticeable differences remain in the positive-field tail.

To assess quantitatively how closely the degraded ViSP and HMI \(B_{\mathrm{los}}\) distributions agree overall, we calculated the mean, median, standard deviation, and skewness. Table~\ref{tab:blos_statistics} summarizes these statistical measures for the degraded ViSP and HMI \(B_{\mathrm{los}}\) distributions. The degraded ViSP and HMI \(B_{\mathrm{los}}\) distributions have mean values of \(6.3\) and \(4.9\) G, median values of \(1.7\) and \(1.3\) G, standard deviations of \(22\) and \(21\) G, and skewness values of \(0.63\) and \(0.50\), respectively. These statistical measures indicate broadly similar distributions, although some differences remain, consistent with the discrepancies visible in their histograms.

The close similarity is expected because both datasets sample the same solar region, and degrading the ViSP data reduces the fine-scale structure and brings its effective spatial resolution closer to that of HMI. However, exact overlap is not expected because the two instruments differ in their effective spatial resolution, instrumental profiles, and observing techniques. The ground-based ViSP observations may also be affected by atmospheric seeing. In addition, HMI is an imaging filtergraph, whereas ViSP is a slit-based spectropolarimeter. These instrumental differences may contribute to the remaining discrepancies between the two \(B_{\mathrm{los}}\) distributions, particularly in the positive-field tail.

In contrast, the broader distribution of the original ViSP \(B_{\mathrm{los}}\) demonstrates that ViSP detects small-scale magnetic features that are partially averaged out in the HMI observations.

\begin{table}[ht]
\caption{Statistical comparison of the degraded ViSP $B_{\mathrm{los}}$ and HMI $B_{\mathrm{los}}$ distributions.\label{tab:blos_statistics}}

\noindent\hspace*{-0.05\textwidth}%
\begin{minipage}{0.50\textwidth}
\centering
\resizebox{\linewidth}{!}{%
\begin{tabular}{|l|c|c|}
\hline
Statistic & Degraded ViSP ME & HMI \\
\hline
Mean [G]               & 6.3  & 4.9  \\
\hline
Median [G]             & 1.7  & 1.3  \\
\hline
Standard deviation [G] & 22   & 21   \\
\hline
Skewness               & 0.63 & 0.50 \\
\hline
\end{tabular}%
}
\end{minipage}
\end{table}

\section{Conclusions}\label{sec:Con}

Over the past few decades, systematic comparisons of $B_{\mathrm{los}}$ measurements have shown that spatial resolution strongly affects inferred photospheric magnetic fields. Earlier Solar and Heliospheric Observatory (SOHO)/MDI, Advanced Stokes Polarimeter (ASP), SDO/HMI, and multi-observatory comparisons showed that weak-field measurements are especially sensitive to instrumental resolution and unresolved mixed-polarity flux cancellation \citep{2003SoPh..213..213B,2012SoPh..279..295L,2014SoPh..289.3483H,2014SoPh..289..769R}. High-resolution SUNRISE/IMaX observations further demonstrated that small-scale photospheric magnetic features can be resolved at sub-arcsecond spatial scales \citep{2010ApJ...723L.164L}. More recently, \citet{2024A&A...686A.218N} compared SDO/HMI magnetograms with high-resolution observations obtained using the CRisp Imaging SpectroPolarimeter (CRISP) at the Swedish 1-m Solar Telescope (SST), demonstrating that small-scale magnetic flux emergence episodes detected by SST/CRISP can remain indiscernible in HMI observations, leading to an underestimation of the associated magnetic flux.
The distinctive aspect of this work is the use of ViSP spectropolarimetric observations, which provide among the highest-resolution measurements of the photospheric magnetic field currently available, to examine a selected field of view within a coronal hole. This high-resolution view allows us to investigate weak, fine-scale magnetic flux that may be spatially averaged or unresolved in HMI observations.

We analyze the $B_{\mathrm{los}}$ derived from ViSP spectropolarimetric observations of the Fe~I~6302~\AA\ line using the ME inversion. We then compare the ViSP ME-derived $B_{\mathrm{los}}$ map with the corresponding HMI Fe~I~6173~\AA\ vector magnetic-field measurement over the common field of view to examine how much small-scale photospheric magnetic flux is resolved by ViSP but remains unresolved in HMI observations. Since both the ViSP and HMI vector magnetic-field maps are obtained through inversion procedures based on ME atmospheric assumptions, this comparison provides a consistent framework for evaluating spatial-resolution effects on the inferred photospheric magnetic field.
The main conclusions of this study are summarized below:
\begin{itemize}
     \item ViSP detects stronger and smaller scale $B_{\mathrm{los}}$ features than HMI within a common coronal hole region. This leads to a larger $\langle |B_{\mathrm{los}}| \rangle$ in ViSP, indicating that HMI systematically underestimates the photospheric magnetic-field strength due to the spatial dilution of small-scale features within larger pixels.

   \item ViSP measures approximately $1.4$ times the total unsigned magnetic flux measured by HMI. This corresponds to HMI measuring approximately $28\%$ less unsigned magnetic flux than ViSP within the common field of view.
   
  \item HMI measures approximately $24\%$ less net signed magnetic flux than ViSP, indicating a difference in the measured dominant-polarity magnetic flux within the observed coronal-hole region.

   \item Despite their different spatial resolutions, ViSP and HMI exhibit remarkably similar magnetic flux imbalance and skewness. The flux imbalance is 65\% for ViSP and 69\% for HMI, while the corresponding skewness values are 0.46 and 0.50, respectively.
 
   \item The observed coronal-hole region remains strongly unipolar in both HMI and ViSP. ViSP reveals additional mixed-polarity magnetic elements at smaller spatial scales; however, these do not change the overall unipolar character of the region.

   \item The ViSP magnetogram provides more detailed photospheric $B_{\mathrm{los}}$ information than the HMI magnetogram. This may improve photospheric boundary conditions for magnetic-field extrapolations. One caveat is the limited ViSP field of view so full coverage of the coronal-hole is needed to determine whether these results apply to the full coronal hole structure.

  \item The degraded-ViSP $B_{\mathrm{los}}$ comparison with HMI $B_{\mathrm{los}}$ shows that the difference between the two measurements is influenced by finite spatial resolution and the associated spatial averaging. When the ViSP map is degraded to HMI-like resolution, its $B_{\mathrm{los}}$ distribution approaches the HMI distribution. Residual differences may also reflect differences in the spectral lines, instrumental characteristics, and magnetic-field retrieval methods used by ViSP and HMI.
  \end{itemize}

Overall, the higher-resolution ViSP measurements recover additional small-scale magnetic flux while preserving the dominant polarity and overall unipolar character of the observed coronal-hole region. Such measurements can provide improved photospheric boundary conditions for magnetic-field extrapolations and for studying the connection between small-scale photospheric fields and the upper solar atmosphere. However, the limited ViSP field of view prevents us from determining the total open magnetic flux of the coronal hole; future observations with broader coronal-hole coverage will be required to do so.
\section{Acknowledgments}\label{sec:acknowledgments}
This research is supported by the DKIST Ambassadors program, funded by the National Solar Observatory through NSF Cooperative Support Agreement AST-1400450. This work is directly related to DKIST observations, and we are grateful for the program's financial support throughout this study. We sincerely thank the anonymous referee for their careful evaluation of our manuscript and for providing insightful and constructive comments and suggestions. Their thoughtful feedback has helped improve the clarity, scientific interpretation, and overall quality of this work.

\section*{Data and Code Availability}

The DKIST data used in this study were obtained
through the DKIST Data Center
(\url{https://dkist.data.nso.edu}).
The SDO/AIA and SDO/HMI data used in this study
were obtained through the Virtual Solar Observatory
(VSO; \url{https://sdac.virtualsolar.org/cgi/search})
and the Joint Science Operations Center
(JSOC; \url{https://jsoc.stanford.edu}).

The three Python computational notebooks developed
for this study are publicly available through Zenodo
\citep{katuwal_2025_16521356,katuwal_2026_23677574,katuwal_2026_23681164}.
These notebooks document the computational procedures
used in the analysis and support the reproducibility
of the results presented in this work.

\section{Appendix}\label{sec:appendix}
\begin{figure*}[!t]
  \centering
  \includegraphics[width=\textwidth]{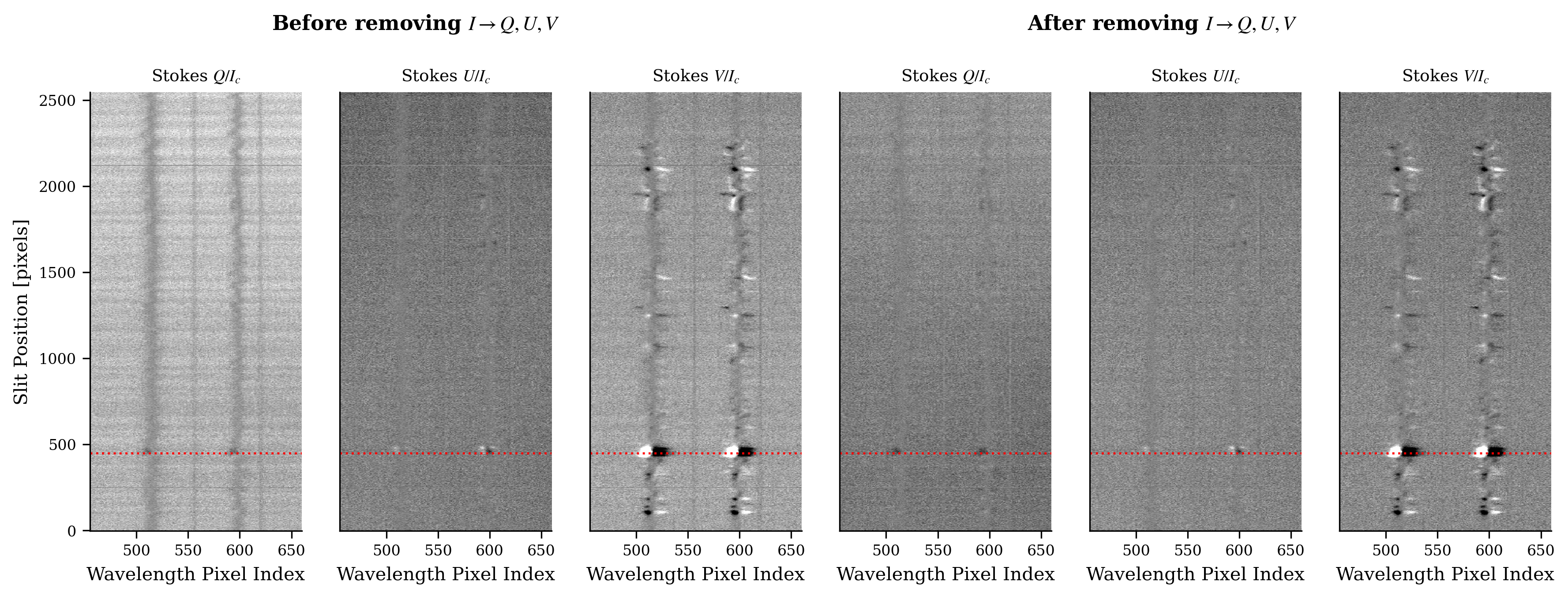}
  \caption{The panels show the Stokes polarization signals before and after removal of the intensity-to-polarization ($I \rightarrow Q,U,V$) crosstalk for one scan step of the ViSP raster observation. The left three panels show the original Stokes $Q/I_c$, $U/I_c$, and $V/I_c$ wavelength--slit maps before correction, while the right three panels display the corresponding maps after applying the crosstalk correction. The horizontal axis represents the wavelength-pixel index, and the vertical axis represents the spatial pixels along the slit. The red dashed line indicates the spatial location of the reference pixel used in the analysis.}
  \label{fig:Crosstalk_correction}
\end{figure*}

The leakage of an unwanted signal from one Stokes parameter into another during polarimetric measurements is known as crosstalk. Crosstalk in the Fe\ \textsc{i}~6302.5\,\AA\ spectra is removed following the method described by ~\citep{1992ApJ...398..359S}, where the leakage of Stokes $I$ into $Q$, $U$, and $V$ is estimated from the continuum and subtracted from the corresponding Stokes profiles. 
Figure~\ref{fig:Crosstalk_correction} illustrates an example of a single raster scan. This specific scan corresponds to raster step 270 and captures the region marked by the blue cross at spatial pixel 450 in both panels of Figure~\ref{fig:FeI6302_Stokes_maps}. The first three panels show the Stokes $Q/I_c$, $U/I_c$, and $V/I_c$ signals before the $I \rightarrow Q,U,V$ crosstalk correction, while the next three panels show the corresponding signals after the correction. The horizontal axis represents the wavelength-pixel index, and the vertical axis represents the spatial pixels along the slit. The red dashed line (same blue cross in Figure~\ref{fig:FeI6302_Stokes_maps}) marks the reference spatial pixel along the slit used to examine the spectra in more detail. In this work, we focus on $B_{\mathrm{los}}$ rather than the full vector magnetic field; however, the complete Stokes information is retained for potential future analysis.

\begin{figure*}[!t]
  \centering
  \includegraphics[width=\textwidth]{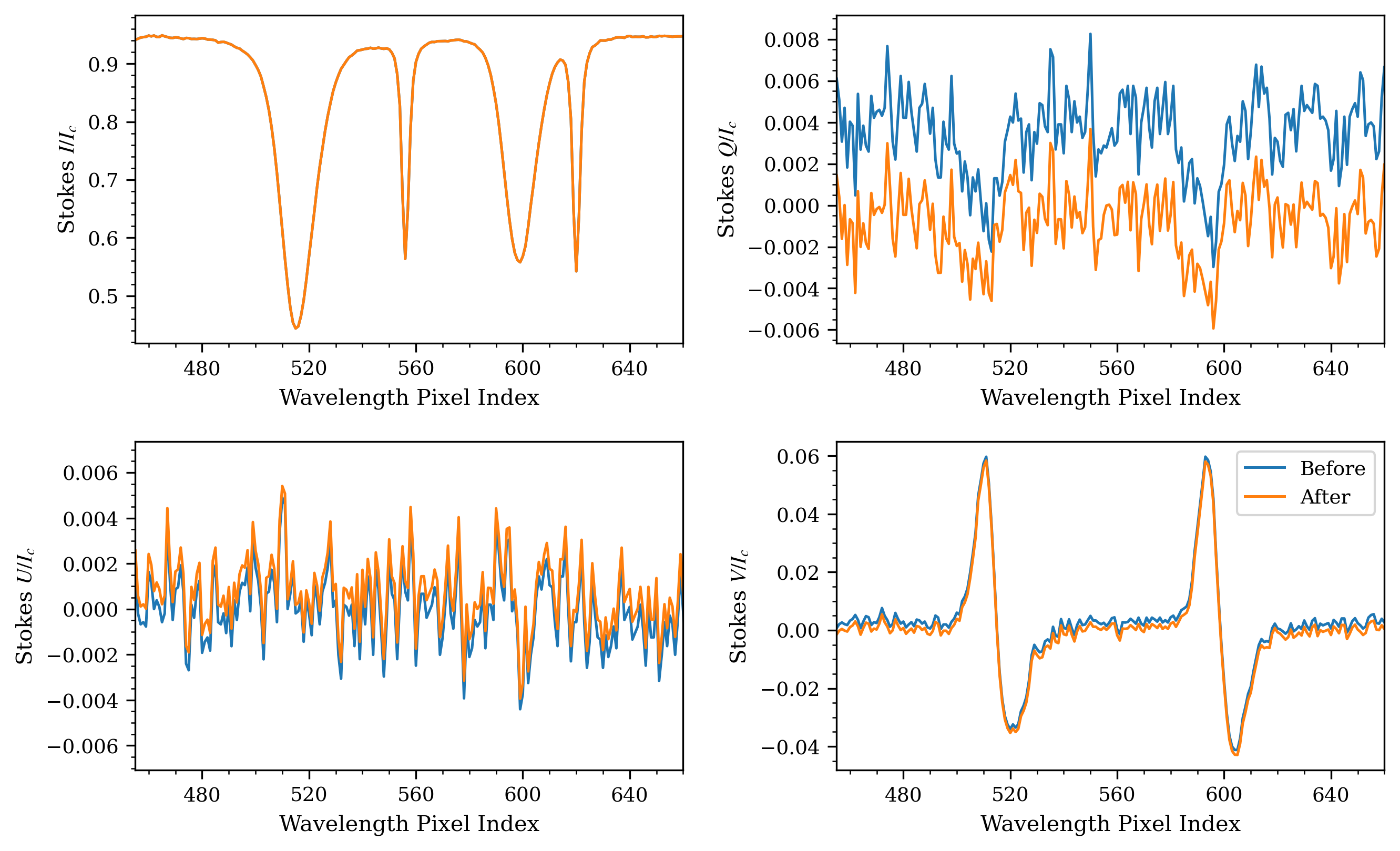}
  \caption{Representative Stokes $I/I_c$, $Q/I_c$, $U/I_c$, and $V/I_c$ spectra at the reference spatial pixel along the slit indicated by the dashed line in Figure~\ref{fig:Crosstalk_correction}. The blue and orange curves show the profiles before and after removal of the intensity-to-polarization ($I \rightarrow Q,U,V$) crosstalk, respectively.}
  \label{fig:single_pixels}
\end{figure*}

To further examine the effect of the crosstalk correction, we extract the Stokes spectra at the reference spatial pixel marked by the red dashed line in Figure~\ref{fig:Crosstalk_correction}. The resulting spectra are shown in Figure~\ref{fig:single_pixels}. The blue curves represent the Stokes profiles before the $I \rightarrow Q,U,V$ crosstalk correction, while the orange curves show the profiles after the correction. The Stokes $I$ profile remains nearly unchanged, while the largest visible modification occurs in Stokes $Q$. In comparison, only relatively small differences are seen in Stokes $U$ and $V$. Since the present study focuses on $B_{\mathrm{los}}$, the comparatively small change in Stokes $V$, which is most directly sensitive to $B_{\mathrm{los}}$, is particularly relevant. To improve the signal-to-noise ratio, we rebinned the spatial pixels and obtained approximately square pixels. The full crosstalk-corrected Stokes $I$,$Q$,$U$,$V$ profiles are subsequently used in the Milne-Eddington inversion. 

\clearpage

\bibliography{Citefile}

\end{document}